# An Analytic Radiative-Convective Model for Planetary Atmospheres

Tyler D. Robinson[1,2,3], and David C. Catling[2,3,4]

(1) Astronomy Department, University of Washington, Box 351580, Seattle, WA 98195-1580, USA;
robinson@astro.washington.edu
(2) NASA Astrobiology Institute's Virtual Planetary Laboratory
(3) University of Washington Astrobiology Program
(4) Department of Earth and Space Sciences, University of Washington, Box 351310, Seattle, WA 98195-1310, USA

## ABSTRACT

We present an analytic 1-D radiative-convective model of the thermal structure of planetary atmospheres. Our model assumes that thermal radiative transfer is gray and can be represented by the two-stream approximation. Model atmospheres are assumed to be in hydrostatic equilibrium, with a power law scaling between the atmospheric pressure and the gray thermal optical depth. The convective portions of our models are taken to follow adiabats that account for condensation of volatiles through a scaling parameter to the dry adiabat. By combining these assumptions, we produce simple, analytic expressions that allow calculations of the atmospheric pressure-temperature profile, as well as expressions for the profiles of thermal radiative flux and convective flux. We explore the general behaviors of our model. These investigations encompass (1) worlds where atmospheric attenuation of sunlight is weak, which we show tend to have relatively high radiative-convective boundaries, (2) worlds with some attenuation of sunlight throughout the atmosphere, which we show can produce either shallow or deep radiative-convective boundaries, depending on the strength of sunlight attenuation, and (3) strongly irradiated giant planets (including Hot Jupiters), where we explore the conditions under which these worlds acquire detached convective regions in their mid-tropospheres. Finally, we validate our model and demonstrate its utility through comparisons to the average observed thermal structure of Venus, Jupiter, and Titan, and by comparing computed flux profiles to more complex models.

*Key words:* convection --- radiation mechanisms: general --- planets and satellites: atmospheres --- planets and satellites: general

1. INTRODUCTION

A fundamental aspect of planetary atmospheres is the vertical thermal structure. A one-dimensional (vertical) model can provide a reasonable estimate of a planet's global-mean temperature profile. Simple models can provide insights into the physics behind the thermal profile of an atmosphere without being obscured by the details of numerical models that sometimes can approach the complexities of real atmospheres. The best simple models are those that incorporate the minimum amount of complexity while still remaining general enough to provide intuitive understanding.

The historical analytic approach to computing atmospheric temperature profiles employs radiative equilibrium only (Chandrasekhar 1960, p.293; Goody & Yung 1989, p. 391-396; Thomas & Stamnes 1999, p. 440-450). Thermal radiation is handled according to the two-stream equations (Schwarzschild 1906), and the calculations typically use the "gray" approximation, meaning that the thermal opacity of the atmosphere is assumed to be independent of wavelength, and is represented with a single, broadband value. Such models have been used to study the ancient (Hart 1978) and modern Earth (Pelkowski et al. 2008), and the stability of atmospheres to convection (Sagan 1969; Weaver & Ramanathan 1995). The temperature profiles from purely radiative models resemble planetary thermal profiles in two broad ways: the temperature falls with height in the deep part of the

atmosphere —the troposphere — and above that, a 'stratosphere' forms, which may contain a temperature inversion, and tends toward an isothermal profile.

McKay et al. (1999) used a radiative model with two shortwave solar channels to study the antigreenhouse effect on Titan and the early Earth. For Titan, the second solar channel improved the fit to the observed stratospheric structure by taking account of stratospheric shortwave absorption. More recently, Hansen (2008) and Guillot (2010) derived similar models for application to Hot Jupiters, and studied variations in emerging radiation and temperature profiles that result from the inhomogeneous distribution of stellar irradiation across a planet.

In general, radiative equilibrium models tend to have regions in their tropospheres where the temperature decrease implies that convection should ensue, which is a process not incorporated into the models but is a part of the essential physics of planetary atmospheres. Convection is common to all planetary tropospheres known in the Solar System (Sanchez-Lavega 2010) and is predicted for exoplanet atmospheres (Seager 2010), so radiative equilibrium models neglect the basic physics of thermal structure. While Sagan (1969) and Weaver and Ramanathan (1995) investigated the conditions under which the temperature profiles generated by gray and windowed-gray radiative equilibrium models will be convectively unstable, these authors did not derive analytic radiative-convective equilibrium models.

A simple radiative-convective model, employed in the limit that the atmosphere is optically thin at thermal wavelengths, joins a convective adiabat to an isothermal stratosphere (Pierrehumbert 2010, p. 169-174). Since this optically thin limit seldom applies to realistic planetary atmospheres, it is more common to join the convective adiabat to a gray radiative equilibrium solution (Goody & Yung 1989, p. 404-407; Nakajima et al. 1992) or a windowed-gray radiative equilibrium solution (Caballero et al. 2008). However, the aforementioned models neglect shortwave attenuation of sunlight by the atmosphere, leading to isothermal stratospheres that fail to represent realistic planetary stratospheres for most planets of the Solar System with atmospheres, including all the giant planets.

Realistic radiative-convective solutions to temperature profiles are commonly computed numerically, for example, using convective adjustment. In this method, the statically unstable profile in the deep layer of the atmosphere that is calculated from radiative equilibrium is fixed to a dry or moist adiabatic lapse rate, which accounts for convection. However, because setting the vertical temperature gradient changes the energy fluxes, the tropopause altitude must be adjusted (and temperatures at all altitudes in the troposphere shifted higher or lower) in numerical iterations until the temperature and upwelling flux are continuous (Manabe & Strickler 1964; Manabe & Wetherald 1967).

Here, we present an analytic radiative-convective model that uses two shortwave channels, thus allowing it to be realistically applied to a wide range of planetary atmospheres. The generality and novelty of our model is demonstrated by applying it to a disparate range of worlds, including Jupiter, Venus and Titan. Given the wealth of new problems posed by exoplanets, development of an analytic model with few parameters is likely to be useful for future application to such worlds, for which only limited data are known.

In this paper, we first derive our analytic model of atmospheric structure for a planetary atmosphere in radiative-convective equilibrium (Section 2). We assume that thermal radiative transfer is gray, and we include two shortwave channels for absorbed solar (or stellar) light so that the model can compute stratospheric temperature inversions. A convective profile is placed at the base of the portion of the atmosphere that is in radiative equilibrium, and the model ensures that both the temperature profile and the upwelling flux profile are continuous across the radiation-convection boundary. In Section 3 we explore the general behaviors of variants of our model, demonstrating its ability to provide clear insights, and including an application to strongly irradiated giant planets, including Hot Jupiters. The utility, validity, and generality of our model are demonstrated by comparing it to previous results, and by applying it to Venus, Jupiter and Titan (Section 4).

## 2. MODEL DERIVATION

In this section, we describe the steps to develop an analytic radiative-convective model for a plane-parallel atmosphere, as follows:

1) We derive a differential equation for the vertical thermal radiative fluxes in a gray atmosphere as a function of optical depth and the temperature at each optical depth.

2) We relate the optical depth to pressure. Physics implies a power law dependence of optical depth on pressure, as others have noted previously (Satoh 2004, p. 372).

3) We define temperature in a (convective) troposphere as a function of pressure (or optical depth using the relation from Step 2). This temperature is used to derive a new expression for the vertical thermal fluxes from Step 1, using a boundary condition of a reference temperature at a reference level, such as the surface of a rocky planet or the 1 bar level in a giant planet's atmosphere.

4) We consider a balance of the net thermal radiative flux with absorbed stellar flux and any internal energy flux (which is important for giant planets) to derive expressions for the temperature and thermal flux profiles in the radiative regime above a troposphere.

5) We derive our radiative-convective model by requiring that the analytic expressions for temperature and upwelling thermal radiative flux are continuous at the join between the convective regime examined in Step 3 and the radiative region evaluated in Step 4.

## 2.1. Gray Thermal Radiative Transfer

The flux profiles of thermal radiation and the atmospheric temperature profile are key parameters in a radiative-convective model, and these quantities are inter-related. The forms of these profiles are different in the regions of the atmosphere that are convection-dominated versus radiation-dominated. We start by writing the general equations that describe the relationships between these key quantities.

In a 1-D, plane-parallel atmosphere, the two-stream Schwarzschild equations for the upwelling and downwelling thermal radiative fluxes ($F^+$ and $F^-$, respectively) are (Andrews 2010, p. 84)

$$\frac{dF^+}{d\tau} = D(F^+ - \pi B) \tag{1}$$

$$\frac{dF^-}{d\tau} = -D(F^- - \pi B) \tag{2}$$

where $\tau$ is the gray infrared optical depth (0 at the top of the atmosphere and increasing towards larger pressures), which is used as a vertical coordinate. For brevity, we will simply refer to these fluxes as "thermal fluxes" for the remainder of this paper. The variable $B$ is the integrated Planck function, with

$$\pi B(\tau) = \sigma T^4(\tau) \tag{3}$$

where $\sigma$ is the Stefan-Boltzmann constant ($5.67 \times 10^{-8}$ W/m²/K⁴) and $T(\tau)$ is the atmospheric temperature profile. The parameter $D$ is the so-called diffusivity factor, which accounts for the integration of the radiances over a hemisphere. The value of $D$ is often taken as 1.66, which compares well with numerical results (Rodgers & Walshaw 1966; Armstrong 1968). Others commonly set $D = 3/2$ (Weaver & Ramanathan 1995), or $D = 2$, which is the hemi-isotropic approximation. For clarity, we note that others (e.g., Andrews (2010) or Pierrehumbert (2010)) sometimes absorb the value of $D$ into their definition of $\tau$. We choose to leave it in our expressions so that the diffusivity approximation is not hidden. As we shall see in Section (2.3), Equations (1) and (2) allow us to solve for the upwelling and downwelling thermal flux profiles if $T(\tau)$ and a boundary condition are specified.

The net thermal flux, $F_{net}$, is given by

$$F_{net} = F^+ - F^- \qquad (4)$$

so that we can combine Equations (1) and (2) to yield

$$\frac{d^2 F_{net}}{d\tau^2} - D^2 F_{net} = -2\pi D \frac{dB}{d\tau} \qquad (5)$$

which is a differential equation that can be used to solve for $B(\tau)$, and thus the atmospheric temperature profile, if $F_{net}(\tau)$ and a boundary condition is known.

*2.2. Relating Optical Depth and Pressure*

While the vertical coordinate of the equations governing the transfer of thermal radiation is optical depth, we must relate this to atmospheric pressure, which is the natural physical vertical coordinate of planetary atmospheres. A vertical pressure coordinate is particularly useful and unifying between atmospheres when we consider lapse rates in tropospheres (Section 2.3).

We take the relation between gray thermal optical depth and atmospheric pressure to be given by a power law in the form

$$\tau = \tau_0 \left( \frac{p}{p_0} \right)^n \qquad (6)$$

where $p$ is atmospheric pressure, $\tau_0$ is the gray infrared optical depth integrated down from the top of the atmosphere to the reference pressure $p_0$, and $n$ is a parameter that controls the strength of the scaling. In the simplest case, when an absorbing gas is well mixed and the opacity does not depend strongly on pressure, we will have $n = 1$. This can physically correspond to Doppler broadening. A common scenario is to have a well-mixed gas providing collision-induced opacity (e.g., $H_2$ in gas giant atmospheres in the Solar System) or pressure-broadened opacity so that $n = 2$. Typically, $n$ takes a value between 1 and 2, as has been parameterized into some radiative models (Heng et al. 2012). In some cases, where the mixing ratio of the absorbing gas depends strongly on pressure, larger values of $n$ have been proposed, such as for water vapor in Earth's lower troposphere (Satoh 2004, p. 373; Frierson et al. 2006).

*2.3. The Convective Regime*

In the convection-dominated region of a planetary atmosphere, we take the temperature-pressure profile to be similar to a dry adiabat, although somewhat less steep because of, for example, the latent heat released by condensation of volatiles during convective uplift. By specifying $T(p)$, we can then derive the expressions for the upwelling and downwelling thermal flux profiles, which we can join to the profiles from the radiatively-dominated region of the atmosphere.

The dry adiabatic temperature variation in the lower, convective part of a troposphere is given by Poisson's adiabatic state equation (e.g., Wallace & Hobbs 2006, p. 78):

$$T = T_0 \left( \frac{p}{p_0} \right)^{(\gamma-1)/\gamma} \tag{7}$$

Here $T_0$ is a reference temperature at a reference pressure $p_0$, and, $\gamma$, is the ratio of specific heats at constant pressure ($c_p$) and volume ($c_v$):

$$\gamma = \frac{c_p}{c_v} \tag{8}$$

Kinetic theory also allows us to relate the ratio of specific heats to the degrees of freedom, $N$, for the primary atmospheric constituent(s), where

$$\gamma = 1 + \frac{2}{N}. \tag{9}$$

Most appreciable atmospheres in the Solar System are dominated by linear diatomic gases, such as $H_2$ (Jupiter, Saturn, Uranus, and Neptune), $N_2$ (Titan), or an $N_2$-$O_2$ mixture (Earth). These molecules have 3 translational and 2 rotational degrees of freedom, so that $N = 5$. Thus, $\gamma = 7/5 = 1.4$ is the same for these worlds. In the case of Venus and Mars, whose atmospheres are dominated by $CO_2$, there are 3 translational, 2 rotational, and ~2 vibrational degrees of freedom (weakly depending on temperature) (Bent 1965), so $N = 7$ and $\gamma = 1.3$. Thus, in Equation (7), the dry adiabatic temperature $T$ varies as $p^{0.3}$ and $p^{0.2}$ for diatomic and $CO_2$ dominated atmospheres, respectively. These $T$-$p$ scalings will also apply to exoplanet tropospheres.

Planetary tropospheres do not follow a true dry adiabat, though, because of the condensation of volatiles during the convection process, such as the effects of water on Earth or methane on Titan, or because the degrees of freedom for the primary atmospheric constituents vary with altitude. Thus, following Sagan (1962), we modify the temperature structure in the convection-dominated region of a planetary atmosphere as

$$T = T_0 \left(\frac{p}{p_0}\right)^{\alpha(\gamma-1)/\gamma} \quad (10)$$

Here, $\alpha$ is a factor, typically around 0.6-0.9, which accounts for small deviations from the dry adiabatic lapse rate, primarily because of latent heat release. Physically, $\alpha$ represents the average ratio of the true lapse rate in the planet's convective region to the dry adiabatic lapse rate. While this parameterized representation of the temperature-pressure profile works well for all of the worlds of the Solar System with thick atmospheres, very moist atmospheres will have temperature varying as $1/\ln(p/p_0)$ according to the Clausius-Clapeyron relation (Satoh 2004, p. 254).

By solving Equation (6) for $p/p_0$ and inserting this into Equation (10), we can rewrite the temperature profile with the gray infrared optical depth as the vertical coordinate

$$T = T_0 \left(\frac{\tau}{\tau_0}\right)^{\alpha(\gamma-1)/n\gamma} = T_0 \left(\frac{\tau}{\tau_0}\right)^{\beta/n} \qquad (11)$$

where we define $\beta = \alpha(\gamma-1)/\gamma$. Inserting this temperature profile into Equation (1) and solving (see Appendix), we obtain an expression for the upwelling thermal flux in the convective portion of the atmosphere (see also Mitchell 2007; Caballero, et al. 2008)

$$F^+(\tau) = \sigma T_0^4 e^{-D(\tau_0-\tau)} + D\sigma T_0^4 \int_\tau^{\tau_0} \left(\frac{\tau'}{\tau_0}\right)^{4\beta/n} e^{-D(\tau'-\tau)} d\tau' \qquad (12)$$

where we have used the boundary condition $F^+(\tau_0) = \sigma T_0^4$, which assumes that the reference location in the atmosphere is either a solid surface or sufficiently deep so that the opacity is large enough to drive the upwelling and downwelling thermal fluxes towards that of a blackbody radiating at the reference temperature. Using the upper incomplete gamma function $\Gamma$ (defined in the Appendix), we can write Equation (12) as

$$F^+(\tau) = \sigma T_0^4 e^{D\tau} \left[ e^{-D\tau_0} + \frac{1}{(D\tau_0)^{4\beta/n}} \left( \Gamma\left(1+\frac{4\beta}{n}, D\tau\right) - \Gamma\left(1+\frac{4\beta}{n}, D\tau_0\right) \right) \right] \qquad (13)$$

which provides an analytic expression for the upwelling thermal flux in the convective portion of the atmosphere. The incomplete gamma function ought to be considered no different from a sine or cosine, in the sense that it is evaluated with a single function in modern programming languages, such as MATLAB, IDL, or Python, so that the flux can be computed readily, if all input parameters are specified. Note that Equation (12) and Equation (13) only apply in the region $\tau_{rc} \leq \tau \leq \tau_0$, where $\tau_{rc}$ is the optical depth at the boundary between the convection-dominated region and the radiation-dominated region. Also, in deriving these expressions, we have assumed that $n$, $\gamma$, $\alpha$ are constant parameters through the convection-dominated region. Physically, the first expanded term, $\sigma T_0^4 e^{-D(\tau_0 - \tau)}$, in Equation (13) is the upwardly attenuated contribution to the upwelling thermal flux from the reference level at temperature $T_0$. The second expanded term is the flux contribution from the atmosphere above the reference level.

A similar line of argument (see Appendix) allows us to write an expression for the downwelling thermal flux in the convective region of the atmosphere

$$F^-(\tau) = F^-(\tau_{rc}) e^{-D(\tau - \tau_{rc})} + D\sigma T_0^4 \int_{\tau_{rc}}^{\tau} \left(\frac{\tau'}{\tau_0}\right)^{4\beta/n} e^{-D(\tau - \tau')} d\tau' \qquad (14)$$

where the boundary condition is that the downwelling flux at the top of the convection-dominated region is equal to the downwelling flux coming from the radiation-dominated region above. While this equation is not required in order to obtain the solution to the radiative-convective thermal structure, it does offer a means to compute the downwelling thermal flux in the convective region.

*2.4. The Radiative Regime*

Above the convection-dominated region of the planet's atmosphere, the temperature profile tends towards radiative equilibrium. In this region, emitted thermal radiation, absorbed stellar radiation, and any source of energy from the planet's interior are all in balance. Thus, by writing an expression for the profile of absorbed stellar flux, we can solve for the thermal flux profiles and the temperature profile using the results from Section 2.1.

The planetary atmospheres of the Solar System exhibit absorption of solar radiation deep in the atmosphere (in the troposphere or at the surface) as well as in the stratosphere. The latter can lead to an inversion in the atmospheric temperature profile, while the former can cause the upper part of the troposphere to be radiatively-dominated to considerable depth, as in the case of Titan. Consequently, for generality it is necessary to distribute the stellar flux across two shortwave channels, and write the net absorbed stellar flux, $F_{net}^{\odot}$, as

$$F_{net}^{\odot}(\tau) = F_1^{\odot} e^{-k_1 \tau} + F_2^{\odot} e^{-k_2 \tau} \qquad (15)$$

where $F_1^{\odot}$ and $F_2^{\odot}$ are the top-of-atmosphere net absorbed stellar fluxes in the two channels, and $k_1$ and $k_2$ parameterize the strength of attenuation in these two channels, and are a ratio of the visible optical depth to the gray infrared optical depth. (Note that $k_1$ and $k_2$ effectively incorporate a spatial and time mean zenith angle.) This description of the absorbed stellar flux is similar to that of McKay, et al. (1999) except that they only allowed for the attenuation of sunlight in one of their channels. At the top of the atmosphere (i.e., where $\tau = 0$), the net absorbed stellar flux must obey

$$F_{net}^{\odot}(0) = (1-A)\frac{F^{\odot}}{4} \tag{16}$$

where $A$ is the planet's Bond albedo, $F^{\odot}$ is the stellar flux incident on the top of the planet's atmosphere, and the factor of four accounts for an averaging over both the day and night sides as well as over the illuminated hemisphere.

A temperature profile results from a balance of fluxes subject to boundary conditions. Balancing the net thermal flux with the absorbed stellar flux and any internal energy source requires that

$$F_{net}(\tau) = F_{net}^{\odot}(\tau) + F_i \tag{17}$$

where $F_i$ is the energy flux from the planet's interior (e.g., ~5 W/m² for Jupiter (Hanel et al. 1981)), which is assumed to be independent of pressure. By combining Equations (5), (15), and (17), and using the boundary condition that that there is no downwelling thermal radiation at the top of the atmosphere (i.e., where $\tau = 0$), we obtain the temperature profile in the radiation-dominated region (i.e., where $0 \leq \tau \leq \tau_{rc}$)

$$\sigma T^4(\tau) = \frac{F_1^{\odot}}{2}\left[1 + \frac{D}{k_1} + \left(\frac{k_1}{D} - \frac{D}{k_1}\right)e^{-k_1\tau}\right] + \frac{F_2^{\odot}}{2}\left[1 + \frac{D}{k_2} + \left(\frac{k_2}{D} - \frac{D}{k_2}\right)e^{-k_2\tau}\right] + \frac{F_i}{2}(1 + D\tau). \tag{18}$$

The upwelling and downwelling thermal flux profiles in the region where $0 \leq \tau \leq \tau_{rc}$ are then given by

$$F^+(\tau) = \frac{F_1^\odot}{2}\left[1 + \frac{D}{k_1} + \left(1 - \frac{D}{k_1}\right)e^{-k_1\tau}\right] + \frac{F_2^\odot}{2}\left[1 + \frac{D}{k_2} + \left(1 - \frac{D}{k_2}\right)e^{-k_2\tau}\right] + \frac{F_i}{2}(2 + D\tau) \quad (19)$$

$$F^-(\tau) = \frac{F_1^\odot}{2}\left[1 + \frac{D}{k_1} - \left(1 + \frac{D}{k_1}\right)e^{-k_1\tau}\right] + \frac{F_2^\odot}{2}\left[1 + \frac{D}{k_2} - \left(1 + \frac{D}{k_2}\right)e^{-k_2\tau}\right] + \frac{F_i}{2}D\tau. \quad (20)$$

*2.5. The Radiative-Convective Model*

The formulae in the preceding sections describe our simple, 1-D radiative-convective climate model. In applying the model, 10 parameters are often fixed ($p_0$, $T_0$, $n$, $\gamma$, $\alpha$, $F_1^\odot$, $F_2^\odot$, $F_i$, $k_1$, and $k_2$), which leaves two variables, $\tau_0$ and $\tau_{rc}$, that the model provides in a solution. However, if the total optical depth $\tau_0$ is specified then one of the aforementioned parameters (typically $T_0$) can become a variable. The temperature, upwelling flux, and downwelling flux profiles follow Equations (18), (19), and (20), respectively, in the radiative region from the top of the atmosphere down to the optical depth at the radiative-convective boundary, $\tau_{rc}$. For optical depths (or pressures) below this level in the atmosphere, the temperature profile follows the adiabat described by Equation (11), and the upwelling and downwelling flux profiles follow Equations (13) and (14), respectively.

The requirement that the temperature and upwelling flux profiles be continuous at the radiation-convection boundary will implicitly solve for two variables in our model (the downwelling flux profile is guaranteed to be continuous due to the boundary condition applied in Equation (14)). Thus, at $\tau_{rc}$, the upward thermal fluxes given by Equation (13) (for the convective formulation) and Equation (19) (for the radiative formulation) must be equal. We also require the convective temperature (Equation (11)) and radiative temperature (Equation (18)) to be equal, so that

$$\sigma T_0^4 \left( \frac{\tau_{rc}}{\tau_0} \right)^{4\beta/n} = \frac{F_1^\odot}{2} \left[ 1 + \frac{D}{k_1} + \left( \frac{k_1}{D} - \frac{D}{k_1} \right) e^{-k_1 \tau_{rc}} \right] + \frac{F_2^\odot}{2} \left[ 1 + \frac{D}{k_2} + \left( \frac{k_2}{D} - \frac{D}{k_2} \right) e^{-k_2 \tau_{rc}} \right] + \frac{F_i}{2} \left( 1 + D\tau_{rc} \right) . \quad (21)$$

These two equalities result in analytic equations that are transcendental and so must be solved numerically, for example, using Newton's scheme. Model parameters that remain after applying the requirements above must be specified or fit. Greater simplicity may be justified by the transparency of an atmosphere, so that one of the stellar channels can be removed (eliminating $F_2^\odot$ and $k_2$). Also, an internal energy source is irrelevant for many worlds (e.g., Earth, Titan, and Venus), setting $F_i = 0$ in many cases. As mentioned before, the value of $\gamma$ is set by the degrees of freedom associated with the primary atmospheric constituent, and $\alpha$ is around 0.6-0.9 for worlds in the Solar System.

Using the thermal fluxes from Equations (13) and (14), as well as the parameterized net stellar flux from Equation (15), the convective flux, $F_{conv}$, can be easily computed according to

$$F_{conv}(\tau) = F_i + F_{net}^{\odot}(\tau) - \left(F^+(\tau) - F^-(\tau)\right) \tag{22}$$

in the region $\tau_{rc} \leq \tau \leq \tau_0$.

*2.6. The Simplest Radiative-Convective Model*

It is worth considering the simplest form of our radiative-convective model wherein there is no attenuation of sunlight ($k_1 = k_2 = 0$), so that

$$F_{net}^{\odot} = F_1^{\odot} + F_2^{\odot} = (1-A)\frac{F^{\odot}}{4} \tag{23}$$

at all locations in the atmosphere, which reduces our radiative equilibrium expressions to those of a single shortwave channel. In the radiative region, the temperature and thermal flux profiles then simplify to

$$\sigma T^4(\tau) = \frac{F_{net}^{\odot} + F_i}{2}(1 + D\tau) \tag{24}$$

$$F^+(\tau) = \frac{F_{net}^{\odot} + F_i}{2}(2 + D\tau) \tag{25}$$

$$F^-(\tau) = \frac{F_{net}^{\odot} + F_i}{2} D\tau \tag{26}$$

For dealing with the convective region, the equalities that ensure temperature and upwelling flux continuity now become

$$\sigma T_0^4 \left(\frac{\tau_{rc}}{\tau_0}\right)^{4\beta/n} = \frac{F_{net}^{\odot} + F_i}{2}(1 + D\tau_{rc}) \tag{27}$$

and

$$\sigma T_0^4 e^{D\tau_{rc}} \left[ e^{-D\tau_0} + \frac{1}{(D\tau_0)^{4\beta/n}} \left( \Gamma\left(1+\frac{4\beta}{n}, D\tau_{rc}\right) - \Gamma\left(1+\frac{4\beta}{n}, D\tau_0\right) \right) \right]$$
$$= \frac{F_{net}^{\odot} + F_i}{2}(2 + D\tau_{rc})$$
. (28)

The dependence on the Bond albedo, which is incorporated into $F_{net}^{\odot}$, can be removed by scaling the temperatures to the equilibrium temperature, $T_{eq}$, which is defined by

$$\sigma T_{eq}^4 = (1-A)\frac{F^{\odot}}{4} + F_i .$$ (29)

Note that the simple model described in this section is similar to an exercise discussed by Pierrehumbert (2010) (see Problem 4.28, p. 311). However, we (1) generalized the temperature pressure relationship in the convection-dominated region (Equation (10)), (2) generalized the relationship between optical depth and pressure (Equation (6)), and (3) use the upper incomplete gamma function to provide an analytic evaluation of the upwelling thermal radiative fluxes in the convection-dominated region.

## 3. GENERAL PROPERTIES OF THE MODEL

The general behavior of the model described in the previous section provides insights into phenomenon in planetary atmospheres. In this section, we will explore the behavior of (1) our simplest radiative-convective model, (2) a model with a single attenuated stellar channel, and (3) a model with a single attenuated stellar channel and an internal energy source. Later (in Section 5), we will see how the properties of atmospheres of certain worlds within the Solar System are explained by some of the general properties discussed here.

### 3.1. Properties of the Simplest Radiative-Convective Model

Our simplest radiative-convective model (Section 2.6) allows us to make straightforward deductions about the location of radiative-convective boundary over a wide range of conditions. We can combine Equations (27) and (28) to yield an expression for $\tau_{rc}$

$$\left(\frac{\tau_0}{\tau_{rc}}\right)^{4\beta/n} e^{-D(\tau_0-\tau_{rc})}\left[1+\frac{e^{D\tau_0}}{(D\tau_0)^{4\beta/n}}\left(\Gamma\left(1+\frac{4\beta}{n},D\tau_{rc}\right)-\Gamma\left(1+\frac{4\beta}{n},D\tau_0\right)\right)\right]=\frac{2+D\tau_{rc}}{1+D\tau_{rc}} \quad (30)$$

which is independent of the net solar flux or the internal energy flux, and only depends on $\tau_0$ and the parameter $4\beta/n$. Figure 1 shows contours of solutions to Equation (30) for $D\tau_{rc}$ over a range of values for $\tau_0$ and $4\beta/n$. Since typical values of $4\beta/n$ are between 0.3-0.5 for worlds in the Solar System (taking $n=2$), the shaded region in this figure demonstrates that $D\tau_{rc}$ will typically be less than unity for realistic values of $4\beta/n$, placing the "radiating level" or "emission level" of $D\tau=1$ in the convective region of the atmosphere. As was discussed in Sagan (1969), worlds with $D\tau_{rc}<1$ will have "deep" tropospheres and "shallow" stratospheres, whereas worlds with $D\tau_{rc}>1$ will have "deep" stratospheres and "shallow" tropospheres.

Figure 1 also shows that, in the limit that $\tau_0 \gg 1$ and $\tau_0 \gg \tau_{rc}$, the value of $\tau_{rc}$ becomes independent of the value of $\tau_0$. In this limit, the upwelling flux at the radiative-convective boundary is no longer sensitive to the radiation coming from the deepest atmospheric layers, and Equation (30) simplifies to

$$\frac{\Gamma\left(1+\frac{4\beta}{n}, D\tau_{rc}\right)}{(D\tau_{rc})^{4\beta/n} e^{-D\tau_{rc}}} = \frac{2+D\tau_{rc}}{1+D\tau_{rc}}, \tag{31}$$

which only depends on the value of $4\beta/n$. Figure 2 shows the solution to Equation (31), computed over a range of values for $4\beta/n$, which, again, shows that, for common values of $4\beta/n$, worlds will have deep troposheres.

Several previous authors (e.g., Sagan 1969; Weaver & Ramanathan 1995) have shown that the requirement for convective instability in a gray radiative equilibrium model without solar attenuation is

$$\frac{d\log T}{d\log p} = \frac{Dn\tau}{4(1+D\tau)} > \frac{\gamma-1}{\gamma} = \left(\frac{d\log T}{d\log p}\right)_{ad} \tag{32}$$

where the subscript "ad" indicates the (dry) adiabatic value. As mentioned earlier, $\gamma$ typically has a value of 1.3-1.4, so that the right hand side of this expression is typically between 0.23-0.29. Sagan (1969) took the solution to Equation (32) as defining the radiative-convective boundary, which is also shown in Figure 2 (taking $\alpha = 1$). Sagan's solution is not the same as a true radiative-convective solution as he only ensures continuity in temperature across the radiative-convective boundary, whereas a realistic radiative-convective model must also maintain upwelling flux continuity across this boundary, which places our radiative-convective boundary more correctly higher in the atmosphere. At large values of $D\tau_{rc}$ Sagan's solution agrees with ours because, in the optically thick limit, the upwelling flux approaches the local blackbody flux, so that temperature continuity and upwelling flux continuity are equivalent. Note, however, that large values of $D\tau_{rc}$ are only achieved for values of $4\beta/n$ that are far larger than values in the Solar System.

*3.2 Properties of a Model with a Single Attenuated Stellar Channel*

The simplest model without a stellar channel is unstable to convection, as shown in Figure 3. Here, the radiative equilibrium temperature profile (labeled as "no atten."), for an atmosphere with $n=2$, has a portion that is unstable to convection (for $\gamma=1.4$). We can increase the generality of our simplest model by allowing for a single stellar channel with attenuation, obtained from Equation (18) by eliminating terms in $F_2^{\odot}$ and $F_i$, and by dropping the subscripts on the remaining stellar channel. Figure 3 demonstrates example temperature profiles (taking $n=2$) for such a model for different values of $k$, which is the ratio of the stellar optical depth in the single channel to the gray thermal optical depth. The logarithmic temperature gradient, or lapse rate, in such a model is

$$\frac{d\log T}{d\log p} = \frac{nk\tau}{4}\left[\frac{(D^2-k^2)e^{-k\tau}}{kD+D^2+(k^2-D^2)e^{-k\tau}}\right]. \tag{33}$$

Note that, for $k>D$, this expression is strictly negative, and, thus, everywhere stable against convection, and the profile has a temperature inversion, as shown in

Fig. 3. This is similar to an argument in Pierrehumbert (2010, p. 212), albeit in different notation. However, for a given $n$ value, it is possible to have a value of $k <  D$ but still be stable against convection, although in this case there is no temperature inversion. For $k = D$, the temperature profile is isothermal, and is stable against convection for all values of $n$.

We investigated this other threshold by examining the maximum value of the lapse rate. Figure 4 shows the maximum value of the lapse rate, which is only a function of $k$ and $n$. This figure demonstrates that, for a given value of $n$ and $\gamma$ (which defines the dry adiabatic lapse rate), there is a threshold value of the relative stellar absorption, represented by $k$, above which the single stellar channel model is everywhere stable against convection. The thick lines in this figure indicate the values of $k$ for which profiles would be unstable to convection as compared to dry adiabats in a $CO_2$-dominated atmosphere and an atmosphere dominated by a diatomic gas. For example, for the $n = 2$ case, the threshold value of $k$ in a $CO_2$-dominated atmosphere is 0.2, and is 0.1 for an atmosphere dominated by a diatomic gas.

In general, the gray infrared optical depth of the radiative-convective boundary in a model with only a single attenuated stellar channel and no internal heat flux depends on $\tau_0$, $k$, and $4\beta/n$, and can be computed from the expression

$$\left(\frac{\tau_0}{\tau_{rc}}\right)^{4\beta/n} e^{-D(\tau_0-\tau_{rc})}\left[1+\frac{e^{D\tau_0}}{(D\tau_0)^{4\beta/n}}\left(\Gamma\left(1+\frac{4\beta}{n},D\tau_{rc}\right)-\Gamma\left(1+\frac{4\beta}{n},D\tau_0\right)\right)\right]$$
$$=\frac{1+D/k+(1-D/k)e^{-k\tau_{rc}}}{1+D/k+(k/D-D/k)e^{-k\tau_{rc}}},\quad (34)$$

which comes from combining the single-channel versions of Equations (18) and (19) with Equations (13) and (11). Note that this expression does not depend on the absorbed stellar flux. Contours of $\tau_{rc}$ as a function of $\tau_0$ and $k$ are shown in Figures 5a and 5b for two different values of $4\beta/n$ (appropriate for a $CO_2$-dominated atmosphere and for an atmosphere dominated by a diatomic gas, respectively).

Figures 5a and 5b demonstrate that, for small values of $k$, we have that $D\tau_{rc} < 1$, which corresponds to a deep troposphere. However, for values of $k$ larger than about 0.1, we have $D\tau_{rc} > 1$, so that the troposphere is shallow because much of the lower atmosphere is stabilized against convection by the absorption of stellar energy throughout the deep atmosphere, as opposed to depositing this energy abruptly at $p_0$, which will happen in the limit of small values of $k$. As we shall see later, cases corresponding to Fig. 5a and 5b are applicable, respectively, to Venus and Titan. For certain combinations of $\tau_0$ and $k$, there exists two values of $\tau_{rc}$ which satisfy Equation (34), but the larger of the two always represents an unphysical solution where the lapse rate in the radiative regime (from Equation (33)) exceeds that for the adiabat in the convective regime at a range of pressures above the radiative-convective boundary.

*3.3 Properties of a Model with a Single Attenuated Stellar Channel and an Internal Energy Source*

Gray radiative equilibrium models with a single stellar channel and an internal energy source have been used to understand certain properties of Hot Jupiters (e.g., Hansen 2008; Guillot 2010). We can derive a similar model by dropping all terms in $F_2^\odot$ in Equation (18) (and by dropping the subscripts on the remaining stellar channel). The resulting radiative temperature profile only depends on $n$, $k$, and the ratio of the absorbed stellar flux to the internal energy flux, $F^\odot / F_i$. Figure 6 shows example temperature profiles for this model for two different values of $k$ where thickened lines show regions unstable to convection. Both models assume strongly irradiated ($F^\odot / F_i = 10^4$) deep atmospheres with $n = 2$. In Figure 6, the deep convectively unstable zone at $p/p_0 > 100$ is caused by the internal flux of the giant planet rather than the stellar flux. In the case of the curve with $k/D = 0.1$, the stellar flux is absorbed around the base of a detached, convectively unstable zone where the atmosphere is optically thick in the infrared, and this gives rise to the separate, detached convective zone.

The lapse rate in the radiative portions of such a model is given by

$$\frac{d\log T}{d\log p} = \frac{nk\tau}{4}\left[\frac{D^2 + \left(F^\odot / F_i\right)\left(D^2 - k^2\right)e^{-k\tau}}{kD(1+D\tau) + \left(F^\odot / F_i\right)\left(kD + D^2 + \left(k^2 - D^2\right)e^{-k\tau}\right)}\right], \qquad (35)$$

which, depending on the values of $n$, $k$, and $F^\odot/F_i$, can either be (1) everywhere stable against convection, (2) only unstable to convection deep in the atmosphere, or (3) unstable to convection in both deep layers and in a detached convective zone, as has been noticed in more complex, numerical models (Fortney et al. 2007).

To investigate the range of parameter space in which detached convective zones can form, we plotted contours of the gray infrared optical depth where Equation (35) first goes unstable to convection (assuming $\gamma = 1.4$, appropriate for an atmosphere dominated by a diatomic gas) for a range of values for $k$ and $F^\odot/F_i$ (Figure 7). We see that, for large values of $F^\odot/F_i$ and for $k/D$ larger than about 0.1, the profile only becomes stable to convection deep in the atmosphere, where the temperature profile is dominated by the internal energy flux.

For larger values of $k/D$, the stellar flux is absorbed higher in the atmosphere where $D\tau < 1$ so that the energy can be radiated to space and there is no detached convective zone. However, for smaller values of $k/D$, the stellar flux is absorbed deeper down where $D\tau > 1$, so that a steep, super-adiabatic temperature gradient can cause a detached convective region. We find that for $k/D$ smaller than about 0.1, a convective zone separate from that caused by the internal heat flux is possible. This limiting value of $k$ can be seen in Figure 4 where the $n=2$ curve crosses the dry adiabatic lapse rate for $\gamma = 1.4$.

The shaded region of Figure 7 shows the portion of parameter space where a detached convective region can form, which has a dependency on the internal heat flux. Note that such a region can only form in our model for $F^\odot / F_i$ larger than about 3, which is consistent with the results shown in Fortney, et al. (2007). For a fixed absorbed stellar flux, $F^\odot$, larger internal fluxes from a giant planet cause the temperature at depth to increase and the convective region extends higher and higher upwards until it joins the detached convective zone, making a continuous region of convection.

## 4. MODEL COMPARISONS AND APPLICATIONS

In this section we apply our radiative-convective model to Venus, Jupiter, and Titan, and compare with observations, while in the subsequent section (Section 5), we discuss the physical insights of these applications. Venus is a case where we can apply our simplest radiative-convective model (Section 2.6). With Jupiter, we demonstrate the differences between radiative and radiative-convective equilibrium models with and without solar attenuation, which shows the successive improvements in the model hierarchy. Finally, for Titan we show the improvement of using a radiative-convective model compared to the purely radiative equilibrium models of McKay, et al. (1999). In the Titan model, we also demonstrate the behavior when we allow $n$ (see Equation (6)) to vary with pressure in the radiative portion of the atmosphere. For clarity, we have summarized the various input parameters and computed variables for the models presented in this section in Table I.

*4.1. Venus*

Our simplified radiative-convective model can be applied to Venus, which, due to a lack of a stratospheric inversion combined with an opaque atmosphere (at thermal wavelengths), means that we can take $k_1$ and $k_2$ to be roughly equal to zero. Taking Venus' Bond albedo to be 0.76 (Moroz et al. 1985), the mean solar flux absorbed by Venus is about 160 W/m². Venus' atmosphere is primarily $CO_2$, so we take $\gamma = 1.3$, and comparing the average lapse rate in Venus' lower atmosphere to an average dry adiabat gives $\alpha = 0.8$. We take $T_0$ and $p_0$ to be the surface temperature and pressure (730 K and 92 bar, respectively), set $F_i = 0$, and explore two cases where either $n = 1$ or $n = 2$, which bounds the two extremes of $n$ that we would expect.

Figure 8 shows a comparison between Venus' observed temperature-pressure profile and our model-generated profiles for these cases. Our Venus model solves for $\tau_{rc}$ and $\tau_0$, and finds these to be 1 and 400, respectively, for the $n = 1$ case. These values put the boundary between convection-dominated and radiation-dominated, $p_{rc}$, to be 0.2 bar. For the $n = 2$ case, we find $\tau_{rc} = 0.1$, $\tau_0 = 2 \times 10^5$, and $p_{rc} = 0.07$. Our values for $p_{rc}$ agree with the observed boundary, which is located at 0.1-0.3 bar, and varies with latitude (Tellmann et al. 2009). Performing a correlation analysis on the data and the $n=1$ and $n=2$ models yields a square of the correlation coefficient, $r^2$, of 0.99 for both models, indicating a very good fit of the model to the observed temperature-pressure profile.

In Figure 9, we compare our model-generated thermal fluxes (from our $n=1$ case) to those from a line-by-line radiative transfer model that has seen extensive application to Venus (Crisp 1986, 1989; Meadows & Crisp 1996), the Spectral Mapping Atmospheric Radiative Transfer (SMART) model. The agreement is quite good (considering we are using an analytic gray model), with differences tending to be at or below about 10%. Also shown is the curve representing $\sigma T(p)^4$ for our model-generated temperature-pressure profile, which demonstrates the expected result that our model thermal fluxes approach the local blackbody flux at high pressure where there are large optical depths.

### 4.2. Jupiter

Jupiter provides a case in which we compare a hierarchy of approaches to analytically computing atmospheric temperature-pressure profiles. We apply three different versions of our analytic model to Jupiter: a radiative equilibrium model without solar attenuation, a radiative-convective model without solar attenuation, and a radiative-convective model with solar attenuation. Steps up this chain of models incorporate additional physics and, thus, require larger numbers of parameters.

For our purely radiative equilibrium model without solar attenuation, we use Equation (24) to compute the temperature profile, with a modification to include an internal energy source, taking $A = 0.34$, $F^\circ = 50$ W/m², and $F_i = 5.4$ W/m² (Hanel, et al. 1981). In addition, we take $p_0 = 1.1$ bar as a reference level, $n = 2$, and $\tau_0 = 6$. The 1.1 bar reference level was selected to coincide with a value reported in the observed profile, and the value of $\tau_0$ was computed from the broadband Rosseland mean opacity tables from Freedman et al. (2008) by interpolating to Jupiter's metallicity. A Rosseland mean opacity was chosen (rather than a Planck mean) because it should apply in the optically thick, deep atmosphere where we set $p_0$.

Our radiative-convective model without solar attenuation builds on the previous model. In the convective regime, we specify the ratio of specific heats as $\gamma = 7/5$ and the average ratio of the lapse rate to dry adiabatic lapse rate as $\alpha = 0.85$. We then add solar attenuation to this model, which develops a stratospheric inversion. Using outputs from the model atmospheres of Fortney et al. (2011), we set $F_1 = 1.3$ W/m² (taken as the solar flux absorbed above 0.1 bar) and $F_2 = 7$ W/m² (taken as the solar flux absorbed below 0.1 bar). About 2 W/m² are absorbed between 0.1 bar and 1.1 bar, so that $k_2 = -\ln((7-2)/7)/\tau_0 = 0.06$, and the value of $k_1$ is found to be 100 by comparing the model temperature profile to the known temperature at the top of Jupiter's stratosphere (165 K) (see Equation (18)).

Figure 10 shows the temperature-pressure profiles from these models as compared to a measured profile synthesized from a number of observations (Moses et al. 2005). The radiative equilibrium model with solar attenuation (model "a") fits through the upper troposphere, but is super-adiabatic at deeper levels. This is corrected in the radiative-convective model without solar attenuation (model "b"), but both aforementioned models cannot match the stratospheric temperature profile since they ignore solar absorption. The radiative-convective model with solar attenuation (model "c") more closely matches the observed temperature-pressure profile, even through the stratosphere. This model finds $p_{rc} = 0.25$ bar, while the radiative-convective model without solar attenuation finds $p_{rc} = 0.22$ bar, which can be compared to the ~0.5 bar boundary found in non-gray numerical radiative-convective models (Appleby & Hogan 1984). Performing a correlation analysis on the data and the radiative-convective model with solar attenuation yields a correlation coefficient of $r^2$ = 0.92, indicating a good fit to the observed temperature-pressure profile, despite having only a few parameters in our model.

In Figure 11 we show the flux profiles from our radiative-convective model with solar attenuation. The difference between the net solar flux and the net thermal flux at the top of the atmosphere is equal to the internal energy flux (5.4 W/m$^2$), and, as is the case for our Titan models (see below), the net solar flux profile shows some structure that is related to its parameterization. An important characteristic of our model is that it can easily compute a convective flux, which is also shown in Figure 11.

*4.3. Titan*

McKay, et al. (1999) applied simple, analytic radiative equilibrium models to Titan. They were able to fit Titan's temperature-pressure curve using a model similar to our Equation (18), except that attenuation was ignored in one of their channels (i.e., extinction of sunlight was ignored in Titan's troposphere, setting $k_2 = 0$). These authors took $n = 4/3$, which is intermediate between the dependence in the troposphere ($n = 2$) and high stratosphere ($n = 1$). Other model parameters were determined by comparing the temperatures in the analytic model to the known temperature-pressure curve (Lellouch et al. 1989) at the surface, tropopause, and the top of the atmosphere, giving $\tau_0 = 3$ and $k_1 = 160$.

For our model, we take $p_0 = 1.5$ bar and $T_0 = 94$ K (the known surface pressure and temperature), $\alpha = 0.77$ (the average ratio of the lapse rate in Titan's troposphere to the dry adiabatic lapse rate), and $\gamma = 7/5$. We also take $F_1^\odot = 1.5$ W/m² and $F_2^\odot = 1.1$ W/m², which represent the solar flux absorbed in Titan's stratosphere and troposphere, respectively. The division of these fluxes between the stratosphere and troposphere follows McKay et al. (1991), while the sum of these fluxes matches the net solar flux absorbed by Titan, and is in agreement with Titan's Bond albedo of roughly 0.27 (Neff et al. 1985) and mean insolation of about 3.6 W/m². Since Titan has no appreciable internal heat flux, we set $F_i = 0$.

Of the 1.1 W/m² that is absorbed in the troposphere, only about 0.35 W/m² is absorbed at the surface (McKay, et al. 1991). This gives an optical depth of roughly unity to sunlight in this channel, so that $k_2 \approx 1/\tau_0$, and $k_1$ is then determined by comparing the model temperature profile to the known temperature at the top of Titan's stratosphere (175 K) (see Equation (18)). As was mentioned earlier, the requirement that the temperature and upwelling flux remain continuous across the radiation-convection boundary allows us to fit for $\tau_{rc}$ and $\tau_0$.

Figure 12 shows a comparison between our radiative-convective model and the best-fit model from McKay, et al. (1999). We show two radiative-convective models: one with $n = 4/3$ and the other in which, following the optical depth profile from McKay, et al. (1999) and the parameterizations from Frierson, et al. (2006) and Heng, et al. (2012), the power of the $\tau$ - $p$ relationship varies smoothly from ~2 at the top of the convective region to ~1 at the top of the atmosphere according to

$$\tau(p) = \tau_{rc}\left[ f\frac{p}{p_{rc}} + (1-f)\left(\frac{p}{p_{rc}}\right)^n \right]. \tag{36}$$

Here $f$ controls the extent of the region of the atmosphere that is dominated by collision-induced absorption ($n = 2$), the top of which we place at roughly 0.2 bar.

For our $n = 4/3$ model, we find $\tau_{rc} = 4.8$, $\tau_0 = 5.3$, $k_1 = 120$, and $p_{rc} = 1.4$ bar. Our value of $\tau_0$ is larger than that of McKay, et al. (1999) because we consider attenuation of sunlight in Titan's troposphere, which then requires a larger greenhouse effect (i.e., more infrared-opaque atmosphere) to maintain the same surface temperature. This also causes our value of $k_1$ to be slightly smaller, since it is inversely proportional to $\tau_0$.

Our variable $n$ model finds $\tau_{rc} = 4.0$, $\tau_0 = 5.4$, $k_1 = 120$, and $p_{rc} = 1.3$ bar. Note that our low values for $p_{rc}$ are in very good agreement with results from Titan general circulation models (Hourdin et al. 1995; Flasar 1998). We discuss the shallowness of Titan's convective region below in Sec. 5 in the context of general model behaviors presented in Sec. 3. Both our variable $n$ model and the $n = 4/3$ model produce good fits to the observed temperature-pressure profile, with the variable $n$ model (dashed line line in Figure 12) performing quite well both the lower and upper atmosphere (yielding a correlation coefficient, $r^2$, of 0.99 when compared to the data). Figure 13 shows the thermal fluxes from our variable $n$ model, and compares the net thermal flux from our model to that from a validated Titan radiative transfer model (Tomasko et al. 2008). The agreement is reasonably good, with discrepancies typically less than 30%, and additional structure appearing in our model due to our parameterization of the net solar flux (Equation (15)).

## 5. DISCUSSION

The radiative-convective equilibrium model of atmospheric structure we have derived provides a simple and intuitive method for analytically computing planetary temperature-pressure profiles. The model also provides straightforward expressions for calculating thermal and convective fluxes. The expressions presented in this work allow for a hierarchical approach to computing atmospheric temperature profiles, with the simplest approach being a radiative equilibrium case without attenuation of sunlight, and the most complex approach being a radiative-convective equilibrium case that includes an internal energy source and the attenuation of sunlight in both the upper and lower atmosphere in two distinct channels.

Our two Venus cases, which had values of $4\beta/n$ equal to either 0.7 (for $n=1$) or 0.4 (for $n=2$) can be compared to the results from the general behavior of a model without stellar attenuation in the large $\tau_0$ limit shown in Figure 2. The $n=1$ case yields a radiative-convective boundary with $D\tau_{rc} \approx 2$, placing the emission level ($D\tau = 1$) in the radiative portion of the atmosphere. However, since we expect pressure broadening and collision-induced absorption to strongly influence infrared opacities throughout much of the Venusian atmosphere, the $n=2$ case should be more realistic. In this case, we have $D\tau_{rc} \approx 0.2$, and the emission level is in the convective portion of the atmosphere. As we expect planetary tropospheres to typically have $n=2$ (or larger), deep tropospheres that radiate from their convective regions will be common for worlds where shortwave opacities are much smaller than longwave opacities, as can be seen for the shaded regions in Figures 1 and 2 for worlds with no shortwave opacity. Fig. 5a and 5b shows that strong shortwave absorption can result in shallow tropospheres because it stabilizes a greater vertical extent of the temperature profile.

By applying a variety of models to Jupiter, we demonstrated how radiative equilibrium models produce super adiabatic regions in a planet's troposphere. Our radiative-convective model corrects this unphysical behavior, and produces a good match to Jupiter's observed temperature-pressure profile. This is especially true of our model that includes attenuation of sunlight in Jupiter's upper atmosphere. This model used $k_1 = 100$, creating a stratospheric inversion where the temperature is decreasing with increasing pressure, as per the discussion in Section 3.2 for single stellar channel models with $k > D$. The tropospheric channel in this model has $k_2 = 0.06$, or $k/D = 0.06/D = 0.04$, so that we expect it to become convectively unstable in comparing to the general model behavior of Figure 4. This model found $D\tau_{rc} = 0.5$, so that the emission level is in the convective regime, and which is consistent with our generalized discussion of a single stellar channel with $k/D = 0.04$.

In applying our radiative-convective model to Titan, we were successful in reproducing the measured temperature-pressure profile and the known result that Titan's atmosphere is only convective in the lowest portions of its "troposphere", from the surface at 1.5 bar to about 1.3-1.4 bar. This shallow convective region demonstrates the awkwardness of applying terrestrial nomenclature to other planetary atmospheres. The term "troposphere" (from the Greek *tropos* for turning, or convective) is universally used to describe layer below the temperature minimum at ~50 km altitude on Titan, but most of this troposphere is radiative, not convective. The reason is because of the absorption of sunlight in Titan's deep atmosphere makes the air very stable against convection.

The tendency for the absorption of sunlight to make Titan's lower atmosphere largely stable to convection is seen in the relatively large value of $k_2$, which we found to be 0.2. Recalling our discussion of a single stellar channel model (Section 3.2), we note that, with this value of $k$ and with $\tau_0 \sim 5$ (as is appropriate for Titan), we expect a shallow troposphere and a deep stratosphere (see Figure 5b with $k/D \approx 0.2/1.66 = 0.12$ ).

In general, our analytic radiative-convective model is a useful tool that can be used to provide insight into behaviors appearing in more complex models and to explain observed phenomenon in planetary atmospheres. Future work could include adding an infrared window to our derivations (see e.g., Parmentier et al. 2012), thus creating a windowed-gray radiative-convective model. Another potential application would be to incorporate our model into retrieval schemes that are currently being developed for exoplanets (Benneke & Seager 2012; Line et al. 2012). As our model incorporates the essential physics of a 1-D planetary atmosphere with only a small number of free parameters, it is well suited to extracting information from exoplanet observations, which are typically sparse.

## 5. CONCLUSIONS

We have derived a simple, 1-D analytic radiative-convective equilibrium model for planetary atmospheric structure. The expressions presented in this work allow for the straightforward calculation of a planet's pressure-temperature profile, thermal flux profiles, and convective flux profile. We have demonstrated the ability of our model to span a wide range of complexity, with the simplest form of our model being fully described by six parameters and two variables (which are solved for implicitly). The model has been successfully applied to a wide range of planetary environments, including Venus, Jupiter, and Titan. IDL (Interactive Data Language) implementations of the analytic model are freely available upon request from the authors for research or pedagogical purposes.

## ACKNOWLEDGEMENTS

This work was performed as part of the NASA Astrobiology Institute's Virtual Planetary Laboratory, supported by the National Aeronautics and Space Administration through the NASA Astrobiology Institute under solicitation No. NNH05ZDA001C. We thank David Crisp, Jonathan Fortney, Tristan Guillot, Vivien Parmentier, and the late Conway Leovy for discussions and insights provided during this project. We also thank Ray Pierrehumbert for an insightful and friendly review.

APPENDIX

The differential equation that describes the upwelling thermal radiative flux in the convective region is obtained by inserting Equation (11) into Equation (1)

$$\frac{dF^+}{d\tau'} = D\left(F^+ - \sigma T_0^4 \left(\frac{\tau'}{\tau_0}\right)^{4\alpha(\gamma-1)/n\gamma}\right). \tag{A1}$$

Rearranging this expression, and defining $\beta = \alpha(\gamma-1)/\gamma$, gives us

$$\frac{dF^+}{d\tau'} - DF^+ = -D\sigma T_0^4 \left(\frac{\tau'}{\tau_0}\right)^{4\beta/n}. \tag{A2}$$

Multiply both sides of this expression by an integrating factor of $e^{-D\tau'}$ gives

$$e^{-D\tau'}\frac{dF^+}{d\tau'} - DF^+ e^{-D\tau'} = \frac{d(F^+ e^{-D\tau'})}{d\tau'} = -D\sigma T_0^4 e^{-D\tau'}\left(\frac{\tau'}{\tau_0}\right)^{4\beta/n}. \tag{A3}$$

This gives us the relation

$$\int_{F^+(\tau_0)e^{-D\tau_0}}^{F^+(\tau)e^{-D\tau}} d(F^+ e^{-D\tau'}) = -\int_{\tau_0}^{\tau} D\sigma T_0^4 \left(\frac{\tau'}{\tau_0}\right)^{4\beta/n} e^{-D\tau'} d\tau' \tag{A4}$$

and using the boundary condition that $F^+(\tau_0) = \sigma T_0^4$ gives us

$$F^+(\tau)e^{-D\tau} = \sigma T_0^4 e^{-D\tau_0} + \int_{\tau}^{\tau_0} D\sigma T_0^4 \left(\frac{\tau'}{\tau_0}\right)^{4\beta/n} e^{-D\tau'} d\tau'. \tag{A5}$$

Thus, multiplying by $e^{D\tau}$ the integral form of the expression for the upwelling thermal radiative flux is (see Equation (12))

$$F^+(\tau) = \sigma T_0^4 e^{-D(\tau_0-\tau)} + D\sigma T_0^4 \int_{\tau}^{\tau_0} \left(\frac{\tau'}{\tau_0}\right)^{4\beta/n} e^{-D(\tau'-\tau)} d\tau' \tag{A6}$$

A standard solution of $-\tau^{4\beta/n+1}E_{-4\beta/n}(D\tau)$ applies to an indefinite integral of the form

$$\int e^{-D\tau}\tau^{4\beta/n}d\tau$$

where $E_n(x)$ is the exponential integral defined as

$$E_n(x) \equiv \int_1^\infty t^{-n}e^{-xt}\,dt\,. \tag{A7}$$

Consequently, Equation (A6) has a solution

$$F^+(\tau) = \sigma T_0^4 e^{-D(\tau_0-\tau)} + D\sigma T_0^4 e^{D\tau}\left[\tau\left(\frac{\tau}{\tau_0}\right)^{4\beta/n} E_{-4\beta/n}(D\tau) - \tau_0 E_{-4\beta/n}(D\tau_0)\right] \tag{A8}$$

The exponential integral is related to the upper incomplete gamma function, defined as

$$\Gamma(a,x) \equiv \int_x^\infty t^{a-1}e^{-t}\,dt \tag{A9}$$

by

$$E_n(x) = x^{n-1}\Gamma(1-n,x) \tag{A10}$$

so that we can rewrite Equation (A8) as

$$F^+(\tau) = \sigma T_0^4 e^{-D(\tau_0-\tau)} + \frac{\sigma T_0^4 e^{D\tau}}{(D\tau_0)^{4\beta/n}}\left[\Gamma(1+4\beta/n, D\tau) - \Gamma(1+4\beta/n, D\tau_0)\right] \tag{A11}$$

which gives us Equation (13). Note that Pierrehumbert (2010, p. 201) derives an analytic expression for the upwelling thermal radiative flux in a convective atmosphere under the assumption that the temperature profile follows a dry adiabat and in the limit that $\tau_0 \to \infty$. While this limit is valid for runaway greenhouse studies, it is not generally true for real planetary atmospheres (e.g., the Titan models in this work).

In a similar fashion, we can solve for the downwelling thermal radiative flux by inserting Equation (11) into Equation (2) to give us

$$\frac{dF^-}{d\tau'} = -D\left(F^- - \sigma T_0^4 \left(\frac{\tau'}{\tau_0}\right)^{4\beta/n}\right) \tag{A12}$$

Following the steps above, and multiplying by an integrating factor of $e^{D\tau'}$, gives us

$$e^{D\tau'}\frac{dF^-}{d\tau'} + DF^- e^{D\tau'} = \frac{d(F^- e^{D\tau'})}{d\tau'} = D\sigma T_0^4 e^{D\tau'} \left(\frac{\tau'}{\tau_0}\right)^{4\beta/n} \tag{A13}$$

which gives us the relation

$$\int_{F^-(\tau_{rc})e^{D\tau_{rc}}}^{F^-(\tau)e^{D\tau}} d(F^- e^{D\tau'}) = \int_{\tau_{rc}}^{\tau} D\sigma T_0^4 \left(\frac{\tau'}{\tau_0}\right)^{4\beta/n} e^{D\tau'} d\tau'. \tag{A14}$$

The integral form of the expression for the downwelling thermal radiative flux is then

$$F^-(\tau) = F^-(\tau_{rc})e^{-D(\tau-\tau_{rc})} + D\sigma T_0^4 \int_{\tau_{rc}}^{\tau} \left(\frac{\tau'}{\tau_0}\right)^{4\beta/n} e^{-D(\tau-\tau')} d\tau'. \tag{A15}$$

Rodgers, C. D., & Walshaw, C. D. 1966, Q J Roy Meteor Soc, 92, 67
Sagan, C. 1962, Icarus, 1, 151
---. 1969, Icarus, 10, 290
Sanchez-Lavega, A. 2010, An Introduction to Planetary Atmospheres (Boca Raton, FL: CRC Press/Taylor & Francis)
Satoh, M. 2004, Atmospheric circulation dynamics and general circulation models (Chichester, UK: Praxis Pub.)
Schwarzschild, K. 1906, Math-phys Klasse, 195, 41
Seager, S. 2010, Exoplanet Atmospheres: Physical Processes (Princeton, N.J.: Princeton University Press)
Tellmann, S., Patzold, M., Hausler, B., Bird, M. K., & Tyler, G. L. 2009, J Geophys Res, 114
Thomas, G. E., & Stamnes, K. 1999, Radiative transfer in the atmosphere and ocean (Cambridge ; New York: Cambridge University Press)
Tomasko, M. G., et al. 2008, Planetary and Space Science, 56, 669
Wallace, J. M., & Hobbs, P. V. 2006, Atmospheric Science: An Introductory Survey (2nd ed.; Burlington, MA: Elsevier Academic Press)
Weaver, C. P., & Ramanathan, V. 1995, J Geophys Res, 100, 11585
43

**Table 1.**

Parameters and computed variables for the models described in Section 4.

| World | $p_0$ [bar] | $T_0$ [K] | $n$ | $\gamma$ | $\alpha$ | $4\beta/n$ | $F_1^\odot$ [W/m²] | $F_2^\odot$ [W/m²] | $F_i$ [W/m²] | $k_1$ | $k_2$ | $\tau_{rc}$ | $\tau_0$ |
|---|---|---|---|---|---|---|---|---|---|---|---|---|---|
| Venus | 92 | 730 | 1 | 1.3 | 0.8 | 0.7 | 0 | 160 | 0 | 0 | 0 | 1 | 400 |
|  |  |  | 2 |  |  | 0.4 |  |  |  |  |  | 0.1 | $2 \times 10^5$ |
| Jupiter | 1.1 | 191 | 2 | 1.4 | 0.85 | 0.5 | 0 | 8.3 | 5.4 | 0 | 0 | n/a | 6 |
|  |  | 168 |  |  |  |  | 0 | 8.3 |  | 0 | 0 | 0.3 |  |
|  |  | 165 |  |  |  |  | 1.3 | 7.0 |  | 100 | 0.06 | 0.3 |  |
| Titan | 1.5 | 94 | 4/3 | 1.4 | 0.77 | 0.7 | 1.5 | 1.1 | 0 | 120 | 0.2 | 4.8 | 5.3 |
|  |  |  | var |  |  | 0.4-0.9 |  |  |  |  |  | 4.0 | 5.4 |

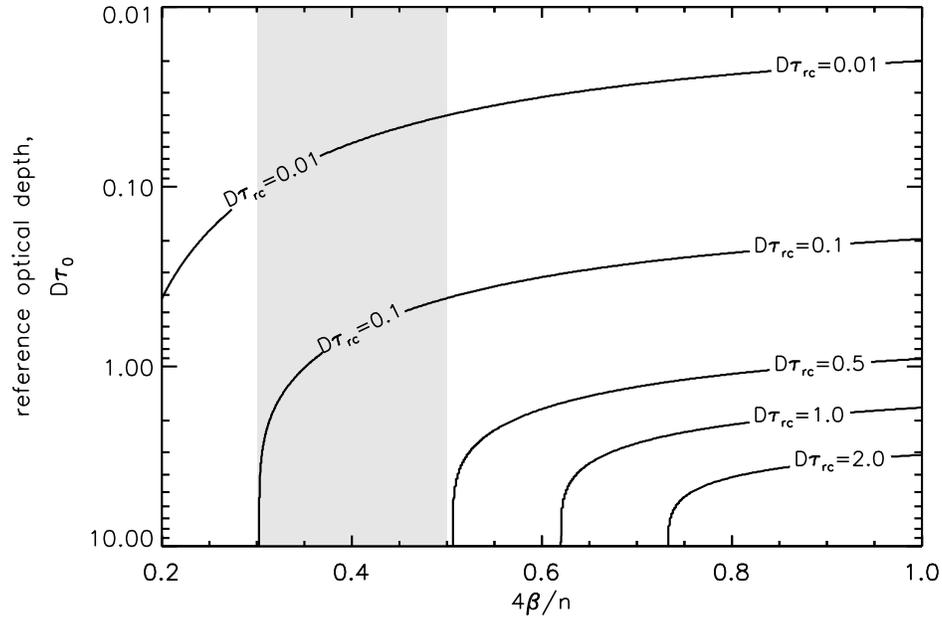

**Figure 1.** Gray infrared optical depth of the radiative-convective boundary ($\tau_{rc}$) for a range of values for $4\beta/n$ and $\tau_0$ for a model without solar attenuation. For the Solar System, the value of $4\beta/n$ is typically 0.3-0.5 (shown as a shaded region), so that we expect deep tropospheres. Note that, for large values of $\tau_0$, the value of $\tau_{rc}$ only depends on $4\beta/n$.

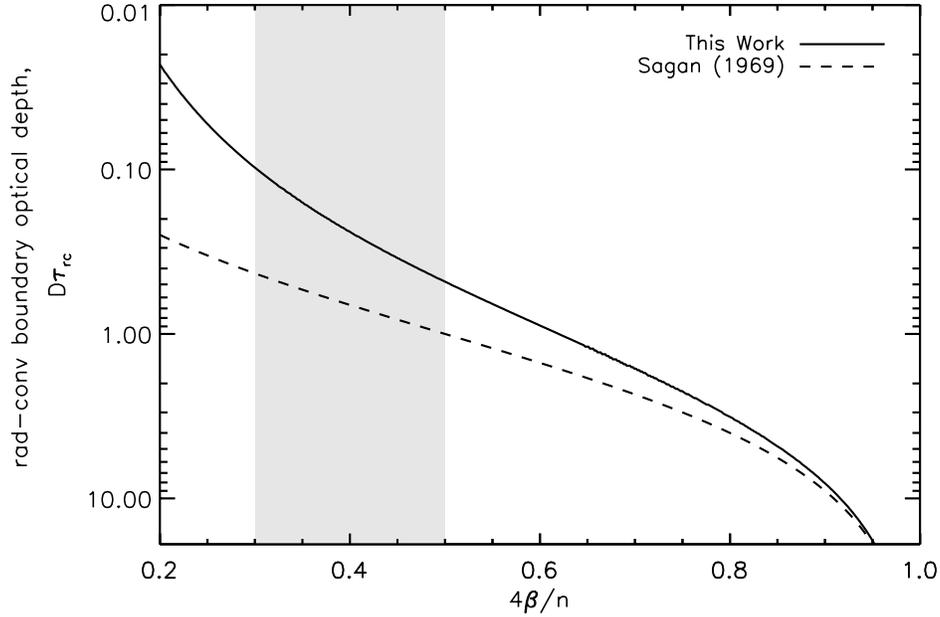

**Figure 2.** The optical depth of the radiative-convective boundary, $\tau_{rc}$, as a function of $4\beta/n$, computed from Equation (31). Note that these values agree with the contours in Figure 2 for large values of $\tau_0$. Also shown are the values of $\tau_{rc}$ from Sagan (1969), who took the radiative-convective boundary to be where the radiative equilibrium profile became unstable to convection. This approach does not guarantee continuity in the upwelling flux profile and, as a result, our radiative-convective boundaries are always at smaller optical depths. The shaded region indicates values of $4\beta/n$ which are typical for the Solar System.

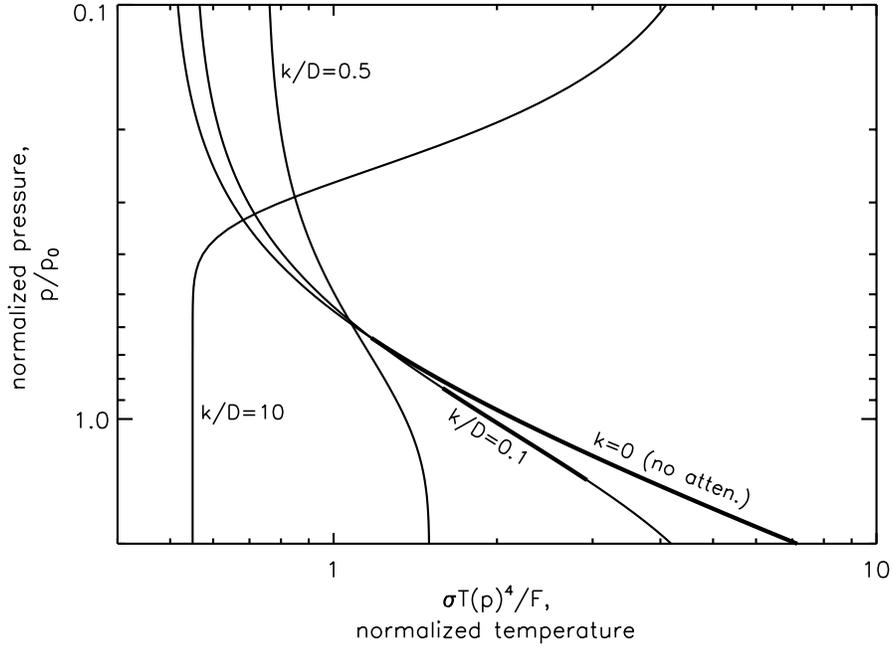

**Figure 3.** Example temperature profiles from our model without shortwave attenuation (labeled "$k=0$ (no atten.)"), and from our model with a single attenuated shortwave channel for several different values of $k$. Pressure has been normalized to $p_0$, and we take $n=2$ and $\tau_0=2$. Temperature, shown as $\sigma T^4$, has been normalized to the sum of the net absorbed stellar flux and the internal energy flux for the model without attenuation, and to the net absorbed stellar flux for the model with attenuation. Thick portions of the curves indicate where the T-p profile is unstable to convection for a dry adiabat with $\gamma=1.4$ (suitable for a world with an atmosphere dominated by a diatomic gas).

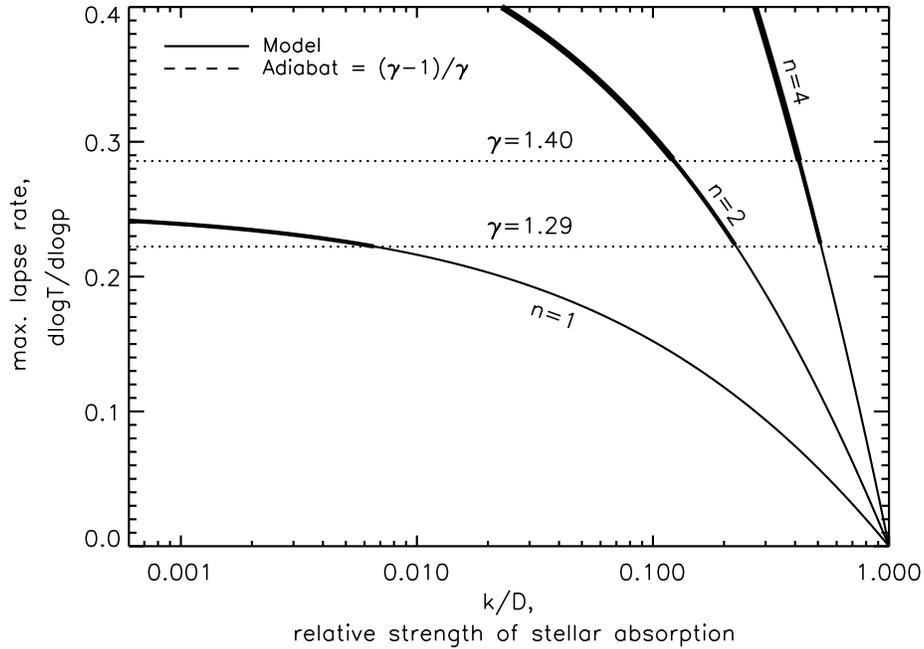

**Figure 4.** The maximum value of the logarithmic temperature gradient from a radiative equilibrium model with a single attenuated stellar channel. Solid lines are for different values of $n$ (Equation (6)), and the horizontal dashed lines are the dry adiabatic lapse rates for $\gamma = 1.29$ and $\gamma = 1.4$, appropriate for a $CO_2$-dominated atmosphere and an atmosphere dominated by a diatomic gas (e.g., Earth and the gas/ice giants), respectively. The thickened portions of the curves indicate the values of $k$ for which models would posses a convectively unstable region for the two aforementioned dry adiabatic lapse rates.

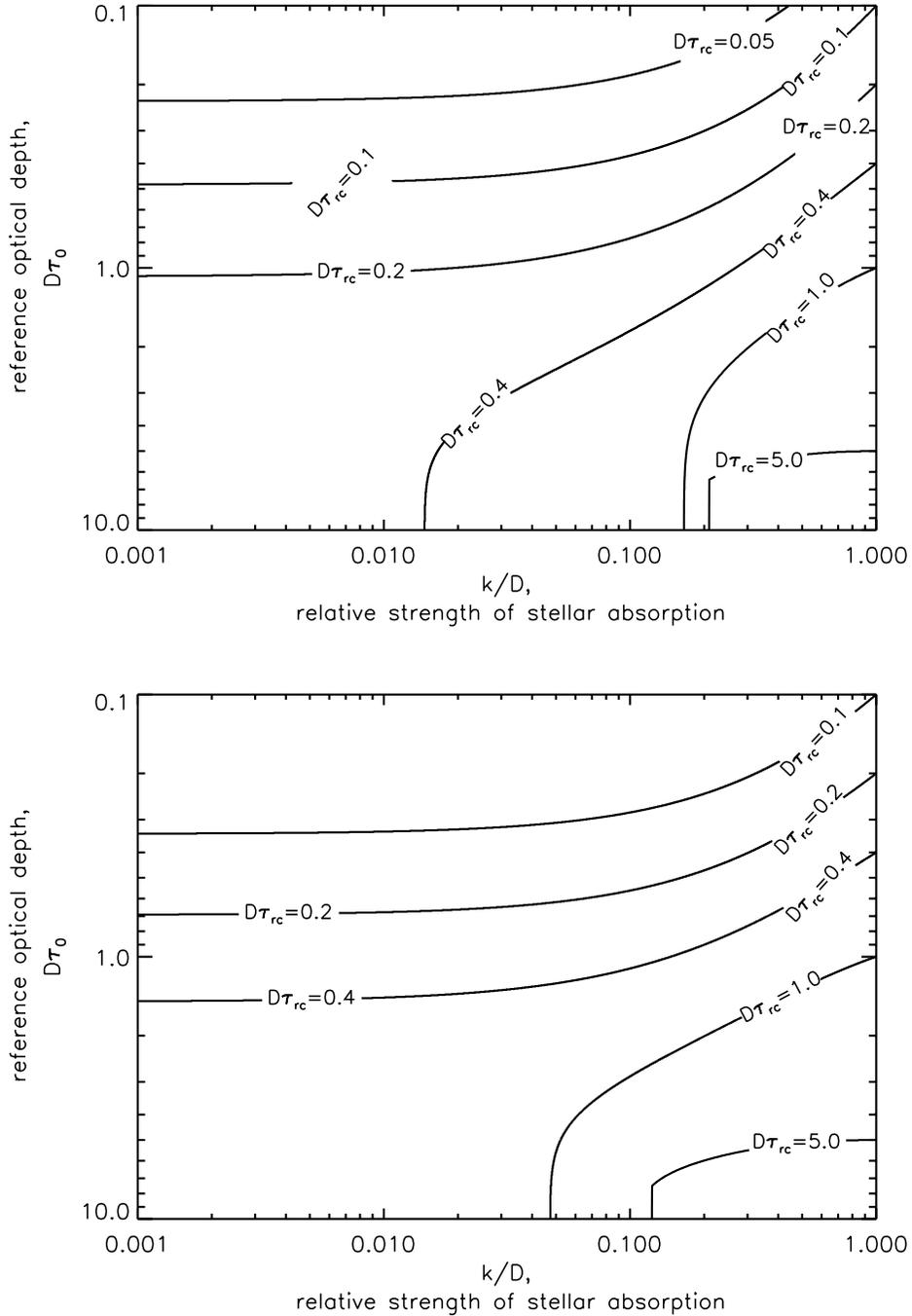

**Figure 5.** Gray infrared optical depth of the radiative-convective boundary, $\tau_{rc}$, for a model with a single attenuated shortwave channel as a function of $k$ and $\tau_0$ for (a) $4\beta/n = 0.46$, which is appropriate for a dry adiabat in a $CO_2$-dominated atmosphere (assuming $n = 2$), and (b) $4\beta/n = 0.57$, which is appropriate for a an atmosphere dominated by a diatomic gas (assuming $n = 2$). In general, increasing $k$ stabilizes deeper portions of the atmosphere against convection, pushing the radiative-convective boundary progressively lower.

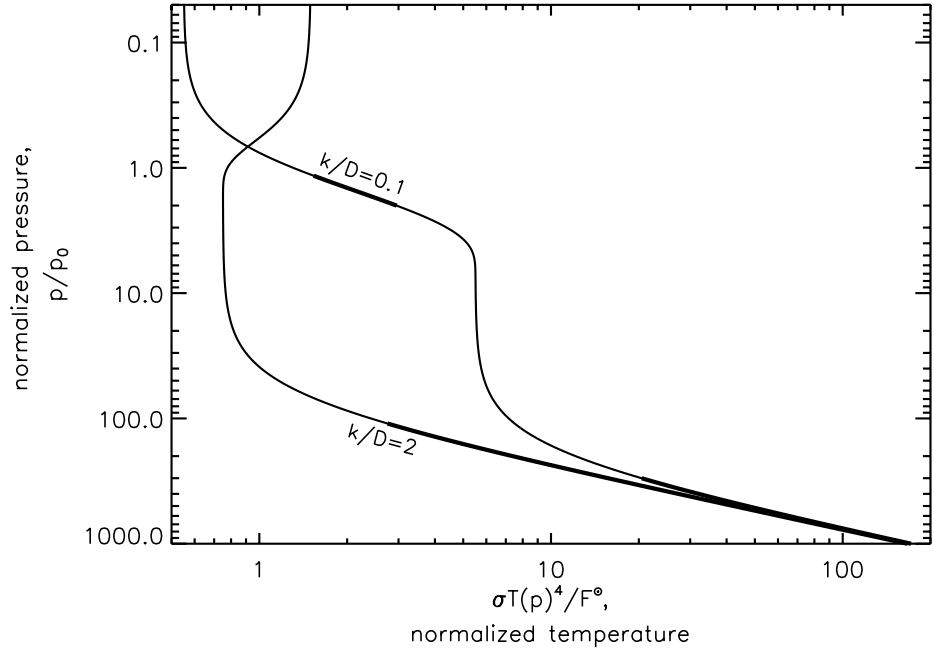

**Figure 6.** Example temperature profiles for a strongly irradiated ($F^\circ / F_i = 10^4$) gas giant for two different values of $k$, and taking $n = 2$ and $\tau_0 = 1$. Pressure has been normalized to $p_0$, and temperature, shown as $\sigma T^4$, has been normalized to the net absorbed stellar flux. Thick portions of the curves indicate where the T-p profile is unstable to convection for a dry adiabat with $\gamma = 1.4$ (suitable for a world with an atmosphere dominated by H$_2$).

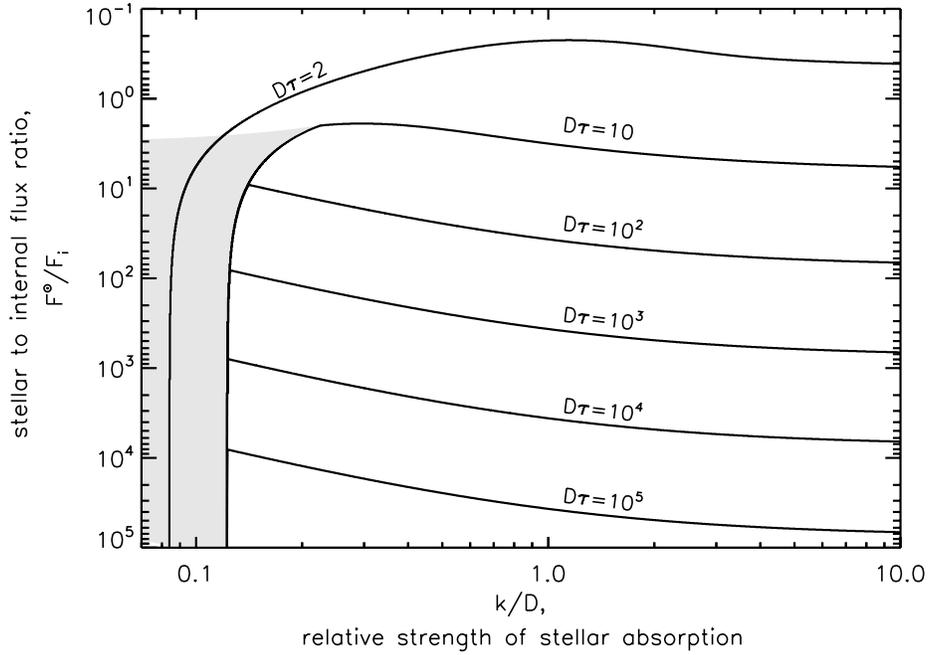

**Figure 7.** Gray infrared optical depth where Equation (35) first goes unstable to convection (for a dry adiabat with $\gamma = 1.4$) for a range of values of $k$ and $F^\odot/F_i$, and taking $n = 2$. This optical depth is large for most values of $k$, but, below $k \approx 0.1$, a detached convective region can form in the mid-troposphere where the lapse rate is dominated by the absorption of sunlight. The shaded region indicates the portion of parameter space where such detached convective regions form.

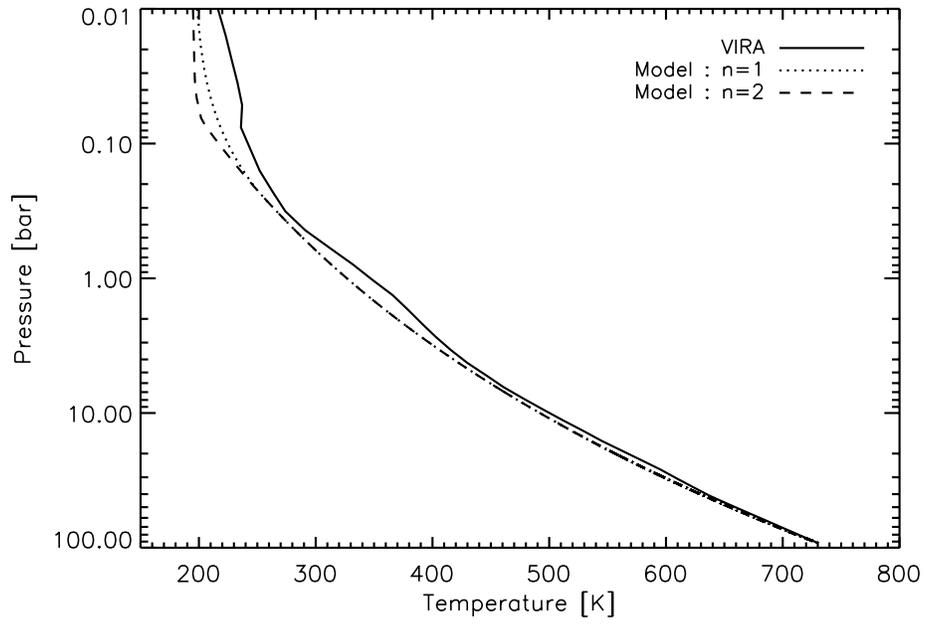

**Figure 8.** Temperature-pressure profiles for Venus. Data is taken from the Venus International Reference Atmosphere (VIRA) (Moroz & Zasova 1997). Models for $n=1$ (dotted) and $n=2$ (dashed) are shown.

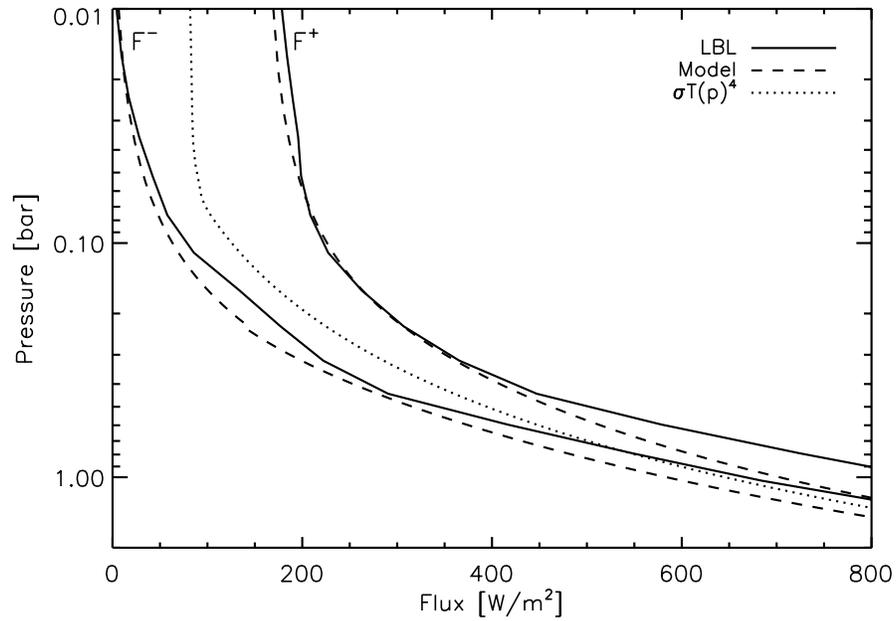

**Figure 9.** Thermal fluxes from our $n=1$ Venus model (dashed) and those from a line-by-line calculation from the line-by-line SMART model (provided by D. Crisp) (solid). Also shown is our model profile corresponding to $\sigma T(p)^4$ (dotted), which our model thermal fluxes approach at large optical depths.

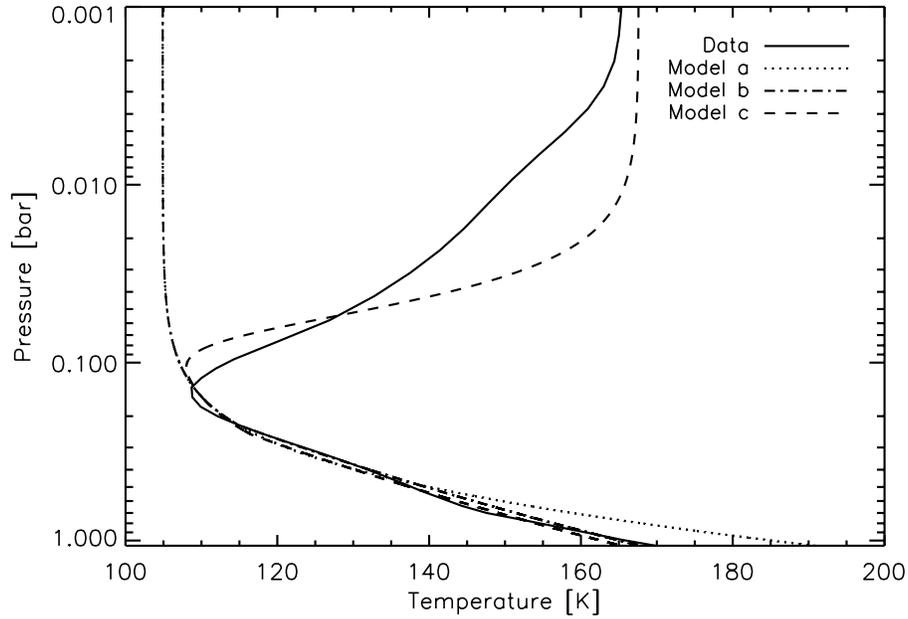

**Figure 10.** Temperature profiles from our hierarchy of Jupiter models and from observations (solid) (Moses, et al. 2005). Model "a" (dotted) is our purely radiative equilibrium model without solar attenuation (which goes super adiabatic in troposphere), model "b" (dash-dot) is our radiative-convective model without solar attenuation, and model "c" (dashed) is our radiative-convective model with solar attenuation.

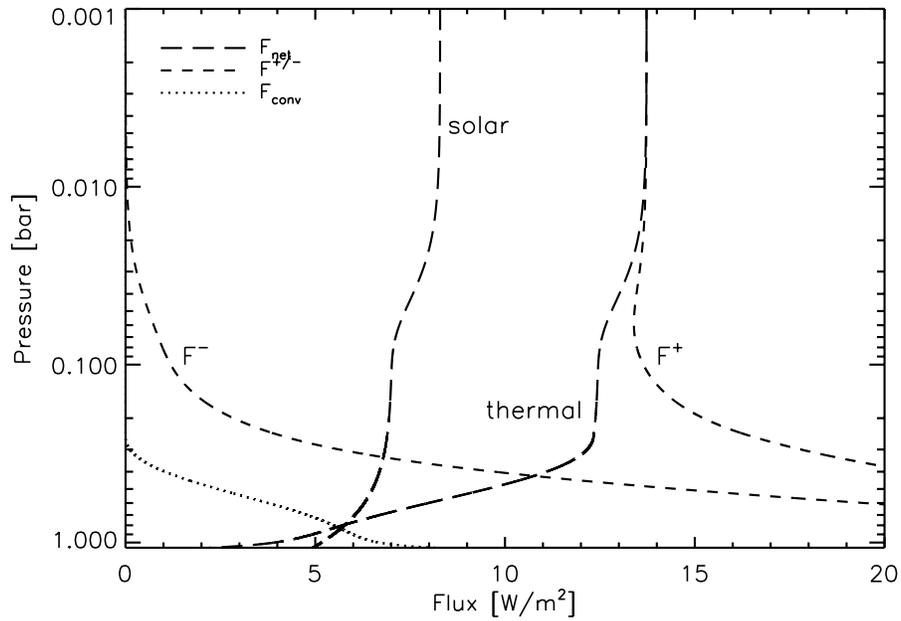

**Figure 11.** Flux profiles from our radiative-convective model of Jupiter that includes solar attenuation. Net solar and thermal fluxes (long dashed) are labeled, and their difference at the top of the atmosphere is equal to the internal energy flux. Also shown are the upwelling and downwelling thermal fluxes (short dashed), and the convective flux (dotted).

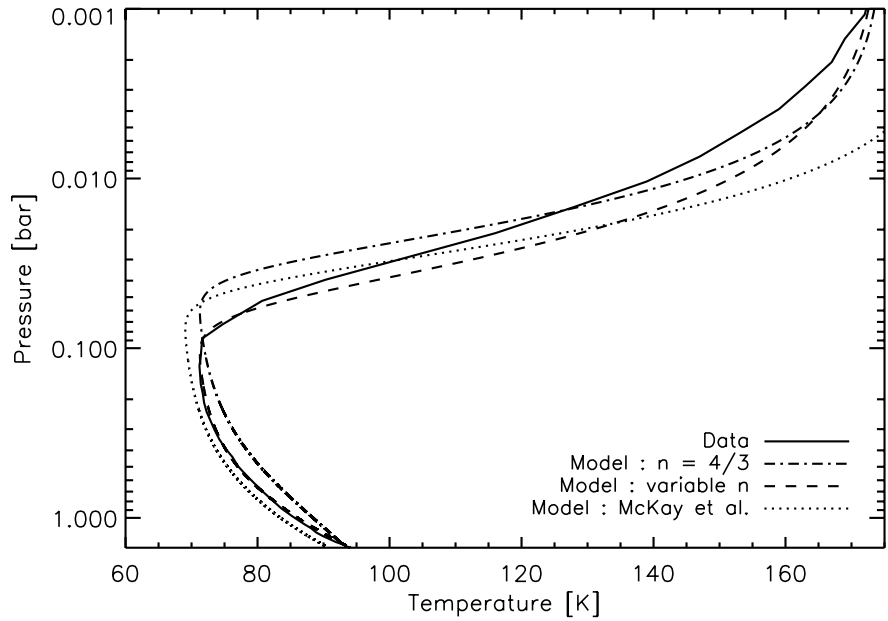

**Figure 12.** Temperature profiles for Titan from a variety of models and from observations (solid line) (Lellouch, et al. 1989). The dotted line is the best-fit analytic radiative equilibrium model from McKay, et al. (1999) which takes $n = 4/3$, the dash-dotted line is for our radiative-convective model which also takes $n = 4/3$, and the dashed line is from our model which allows $n$ to vary smoothly between ~2 near the surface to ~1 in the upper atmosphere.

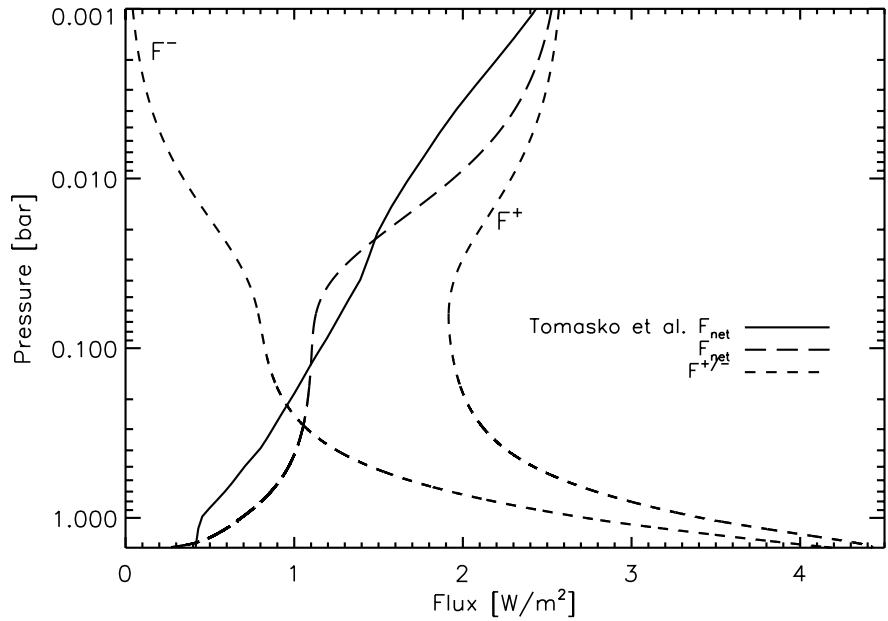

**Figure 13.** Thermal fluxes from our variable $n$ Titan model. Upwelling and downwelling fluxes (short dashed) are labeled, and their difference is the net thermal flux (long dashed). Also shown is the net thermal flux from a Titan radiative transfer model (Tomasko, et al. 2008).